\documentclass[12pt]{article}
\usepackage{epsfig}

\parskip=\medskipamount 
plus.3em minus.5em \abovedisplayshortskip=.5em plus.2em minus.4em
\renewcommand{\thesection}{\arabic{section}}

\catcode`@=11

\@addtoreset{equation}{section} \@addtoreset{equation}{subsection}
\def\theequation{\ifnum\value{section}=0 \arabic{equation}\ignorespaces
\else \ifnum\value{section}=-1 A.\arabic{equation}\ignorespaces
\else \ifnum\value{subsection}=0
\thesection.\arabic{equation}\ignorespaces \else
\thesection.\arabic{subsection}.\arabic{equation}\ignorespaces
                             \fi
                        \fi
                   \fi}

{\catcode`\'=\active \def'{{}^\bgroup\prim@s}}

\catcode`@=12

\newcommand{\bq}{\begin{equation}}
\newcommand{\be}{\begin{equation}}
\newcommand{\fq}{\end{equation}}
\newcommand{\ee}{\end{equation}}
\newcommand{\bqr}{\begin{eqnarray}}
\newcommand{\beqs}{\begin{eqnarray}}
\newcommand{\fqr}{\end{eqnarray}}
\newcommand{\eeqs}{\end{eqnarray}}

\newcommand{\rf}[1]{(\ref{#1})}

\def\bop#1{\setbox0=\hbox{$#1M$}\mkern1.5mu
    \vbox{\hrule height0pt depth.04\ht0
    \hbox{\vrule width.04\ht0 height.9\ht0 \kern.9\ht0
    \vrule width.04\ht0}\hrule height.04\ht0}\mkern1.5mu}

\begin{document}
\thispagestyle{empty}

\begin{flushright}
\begin{tabular}{l}
\end{tabular}
\end{flushright}

\vskip .6in
\begin{center}

{\Large\bf  Fast Factoring of Integers}

\vskip .6in

{\bf Gordon Chalmers}
\\[5mm]

{e-mail: gordon@quartz.shango.com}

\vskip .5in minus .2in

{\bf Abstract}

\end{center}

An algorithm is given to factor an integer with $N$ digits in
$\ln^m N$ steps, with $m$ approximately $4$ or $5$.  Textbook quadratic
sieve methods are exponentially slower.  An improvement with the
aid of an a particular function would provide a further exponential
speedup.

\setcounter{page}{0}
\newpage
\setcounter{footnote}{0}

Factorization of large integers is important to many areas of pure
mathematics and has practical applications in applied math including
cryptography.  This subject has been under intense study for many
years \cite{CrPo}; improvements in the methodology are especially 
desired for computational reasons.

Given an integer $N$ composed of approximately $\ln_{10} N$ digits,
standard textbook quadratic sieve methods generate the factorization
of the number into primes in roughly

\bqr
e^{a\sqrt{\ln N \ln\ln N}}
\label{qssteps}
\fqr
moves.  The steps require manipulations of large integers, of the
size $N$, with bit complexity of approximate $\ln N$.  The number $a$
is approximately $2$, depending on the variant used \cite{CrPo}.

The presentation in this work generates a computational method to
obtain the prime factorization in

\bqr
\ln^m N
\label{improved}
\fqr
moves with integers of the same size.  The factor $m$ is specified
by the convergence of the solution to a set of polynomial equations
in $\ln N$ variables, which numerically is approximately $m=3$, after
the root selection is chosen from small numbers to large (see, e.g. 
\cite{EDM2}).

Given a function $C_N$ that counts the number of prime factors of a
number, i.e.

\bqr
N=\prod_{j=1}^{r} p_{\sigma(j)}^{k_j} \qquad\quad
C_N=\sum_{i=1}^r k_j \ ,
\fqr
the factorization of the number $N$ could be performed in approximately
$C_N^m$ steps.  The bound on the number of prime factors of an integer
$N$ is set by $\ln_2 N$, the product of the smallest prime number $2$.
The number of primes smaller than a number $N$ is approximately $N/\ln N$,
and the $C_N$ is roughly $\ln\ln N$.  Hence, given the function $C_N$
a further exponential improvement is generically given.  However, this
function would drastically simplify the factorization of large numbers
possessing only a few prime factors.  The upper bound of $C_N\sim\ln N$
describes the case discussed in the previous paragraph.

Consider a number with exactly $C_N$ factors.  This number projects in
base $x$ onto the form,

\bqr
N= \sum_{i=0}^{C_N} a_i x^i \ .
\label{baseprojection}
\fqr
The polynomial form in \rf{baseprojection} admits a
product form,

\bqr
N = \prod_{i=1}^{C_N} (c_j x-b_j) \ ,
\label{productprojection}
\fqr
with $c_j x - b_j$ integral.  The number scales into the form, 

\bqr 
N= \gamma \prod_{i=1}^{C_N} (\alpha_j x-1) \ , 
\label{products}
\fqr 
in which there are $C_{N}$ numbers $\alpha_j$ and a number $\gamma$.  
The same integer has the prime factorization

\bqr
N = \prod_{i=1}^{C_N} p_{\sigma(i)} \ ,
\label{primeproduct}
\fqr
with the set $\sigma(j)$ containing possible redundancy, for example,
$p_{\sigma(1)} = p_{\sigma(2)} = 2$.  Given an integer base $x$ the solution
to the numbers $b_j$ generate the prime factors, {\it as long as the value
$C_N$ is correct}.

Two examples are given.  First, $15=3^2 + 2 (3)=x(x+2)$, which solves 
for the prime factors $3$ and $5$.  Second, $10=2^3+2=x(x^2+1)$, which 
solves for the prime factors $2$ and $5$.  In the second example, even 
though there are two factors, the polynomial is a cubic with a vanishing 
zeroth order term; the origin of the cubic is that there is a complex 
root.

The polynomial base form of the number, i.e. $N=\sum a_i x^i$, will 
not in all cases factor into the form \rf{productprojection} with 
real coeffcients $c_j$ and $b_j$.  However, because the coefficients 
are real, the roots will enter in complex conjugate pairs.  The product 
of these complex conjugates form a positive number.  In order to 
test all possible cases, including the presence of a factor being 
represented as a product of two complex roots, all numbers from 
$C_N$ to $2C_N$ should be examined.  Potential complex roots 
come in pairs, and the maximum number of factors could take on the 
form of $C_N$ products of two complex numbers.  The cost of the 
additional complexity is of order unity.    

The expansion of \rf{productprojection} generates a set of algebraic
equations relating the integer coefficients $b_j$ to those in $a_j$.
The form is,
\bqr
\gamma \alpha_1 \alpha_2 \ldots \alpha_{C_N} = a_{C_N}
\cr \ldots \cr 
\gamma = a_{0} \ , 
\fqr
in which the combinations 

\bqr  
c_j x - b_j 
\fqr 
must converge to integers or into pairs with the product being an integer, 
and for the maximal factorization to 
prime numbers $p$ (of which there are an approximate $N/\ln N$ of them 
for the number $N$).  The determination of the numbers $\alpha_j$ must 
be rational as $(c_j/b_j x -1) = n/b_j$, with $n$ integral.  The 
other case of interest is when $c_j/b_j$ is complex and the relevant 
condition is $\vert(c_j/b_j x-1)\vert^2 = n^2/b_j^2$; this is not 
satisified in general by complex rational numbers.  However, one 
may square the number $N$ and then all terms in the product must 
be rational.  

The number $\gamma$ must be an integer or the square of $\gamma$ must be 
an integer, according to the presence of complex terms (roots) which 
square to an integer.  If the solution does not satisfy these 
criteria, then there is not a valid factorization $N$ into integers.  
Given rational solutions $\alpha_j=c_j/b_j$ and the $\gamma=\prod b_j$, 
the straightforward multiplication of $\gamma$ into the $C_{N}$ factors 
generates the factorization into $N_1 N_2 \ldots N_{C_N}$, via eliminating 
the denominators in the individual terms of the rational numbers.  
The complex root case allows the numbers to be determined as $N_j = 
N_{j,+} N_{j,-}$.  

Solving these equations generates the prime factorization of the
integer $N$ into the set of primes $p_{\sigma(j)}$.  Numerically, solving
a set of equations in $n$ variables typically has convergence of $n^3$
if the initial starting values are chosen correctly.

In the case of $C_N$ not known, but bounded by $\ln N$, all cases of
interest from the test cases of ${\tilde C}_N=1$ to ${\tilde C}_N=\ln N$
may be examined, at the cost of duplicating the process by the bound
$\ln N$.  Typical true values of $C_N$ are expected for generic numbers
to be smaller than the bound, e.g. $\ln\ln N$.  The cases from $C_N$ 
to $2C_N$ must also be examined in order to take into account the 
pairs of complex conjugates.  

Computationally exploring all of the cases from ${\tilde C}_N=1$ to
${\tilde C}_N=2\ln N$ (e.g. $\sim C_N$) for integer bases $x$ finds all 
product forms of the integer $N$ into products

\bqr
N = N_1 N_2 \qquad
N = N_1 N_2 N_3
\fqr
\bqr
N = N_1 N_2 \ldots N_{C_{N}} = \prod_{j=1}^{C_N} p_{\sigma(j)} \ .
\fqr
Solving for $b_j$ and $c_j$ (e.g. $\alpha_j=b_j/c_j$) in terms of $a_j$ 
generates either integral 
values or non-integral values, or pairs or complex conjugates.  In the 
case of integral values for all the
$b_j$ parameters, the base $x$ is resubstituted into the factors
$c_j x-b_j$ of the total product,

\bqr
N= \prod^{{\tilde C}_N} (c_j x-b_j) = \gamma \prod (\alpha_j x - 1)\ ,
\fqr
to find the values of the individual $N_n$.  The integral solutions
generate the various factorizations.  In the case of complex conjugate 
pairs, the integrality is tested by multiplying the individual terms, 

\bqr  
(c_jx-b_j)(c_j^* x -b_j^*) \sim (\alpha_j x -1)(\alpha_j^* x - 1) \ , 
\fqr 
and determining if it is an integral (the latter has to be rational).  
The maximum ${\tilde C}_N$ that
results in integral values of $c_j x - b_j$ gives the prime factorization.
Non-integral solutions to $c_j x - b_j$ do not generate integer factorization
of the number $N$ into ${\tilde C}_N$ numbers.  

In the computation for the roots, the parameters $\alpha_j$ and $\gamma$ 
are determined.  The integrality of the $c_j x-b_j$ translates into 
$\alpha_j$ being a rational number (when not a complex root allowing the 
complex $c_j x - b_j$ square to be an integer).  The rationality allows 
the denominators of all the $\alpha_j$ parameters to be extracted and 
used to multiply the prefactor $\gamma$. The case of the complex roots 
may be examined also by first taking the complex square and examining 
for integrality; the denominators must also be taken out of the products.  

Computationally testing all cases from ${\tilde C}_N=1$ to $\ln N$
finds all product forms of the potentially large integer $N$.  The
factorization process entails three steps: 1) projecting the number
$N$ into base $x$, 2) solving the system of algebraic equations for
integral $c_j x - b_j$, 3) substituting the base $x$ back into the 
$c_j x-b_j$ for factor determination.

Computationally the first step requires specifying the base $x$ and
projecting the number onto it.  The base $x$ is specified by the two
equations,

\bqr
\ln x > {{\ln N}\over {\tilde C}_N+1} \qquad
\ln x \leq {{\ln N}\over {\tilde C}_N} \ .
\fqr
Following the determination of $x$, the coefficients $a_j$ are determined
via starting with $a_{{\tilde C}_N}\leq x$, $j={\tilde C}_N$, $N_{{\tilde C}_N}
= N$ and following the procedure,

\bqr
N_{j-1} = N_j - a_j x^j
\fqr
with
\bqr
\gamma_{j-1} = N_{j-1} / x^{j-1} \ .
\fqr
Take $a_{j-1}=[ \gamma_{j-1} ]$ with a rounding down of $\gamma_{j-1}$;
if $a_{j-1}\geq x$ then the procedure stops and the remaining $a_i$ are
set to $a_i=0$ with $i<j-1$.  Otherwise, the subtraction process continues.
The procedure costs at most $3 {\tilde C}_N$ operations with numbers of
at most size $N$ (bit size $\ln N$).  Due to the bound on ${\tilde C}_N$,
this process is at most of the size $3\ln N$ operations, one of which
is division.

The next step requires the solution of the algebraic equations for $b_j$
in terms of $a_j$.  There are ${\tilde C}_N+1$ equations in ${\tilde C}_N+1$
variables.  The initial roots are chosen from the lowest prime neighborhood
around $p=c_j x-b_j$, to larger.  This procedure is natural
for the root determinations in the case of the exact $C_N$.  Another case
is in choosing cascades in the range $10^n$ to $10^{n+1}$ for $n\leq
\ln_{10} N$.  Convergence is not analyzed, but for well chosen starting
values, the number of iterations is typically $N_v^3$; this is
${\tilde C}_N^3$ for the case of ${\tilde C}_N$ variables and equations.
The bound is $\ln^3 N$.  Roughly, if the number of operations per
iteration is ${\tilde C}_N^2$ (i.e. evaluating a set of similar
polynomials) and there are ${\tilde C}_N$ roots, then the steps would
number as ${\tilde C}_N^6$.

If any roots converge to a non-integral value of $b_j$, then the integer
$N$ does not factor into ${\tilde C}_N$ numbers.  This shortens the
number of iterations and steps.  The process of determining the factorization
of $N$ into the products of two to $C_N$ numbers requires $\ln^m N$ steps,
with $m$ denoting an average value from the root selection process, the
number of variables at each step, the root solving and iteration process
including shortcuts such as information from lower ${\tilde C}_N$ examples,
and an averaging of the shortening the algorithm during the process of
lower unknown roots or non-integerness.  Perhaps, the average results in
$m\sim 4$ or $5$, less than $\ln^6 N$.

To compare with the textbook quadratic sieve method, take the logarithm
of the steps for both this method and the former,

\bqr
a\sqrt{\ln N \ln\ln N} \qquad m \ln\ln N
\fqr
which is,

\bqr
a^2 \ln N \ln\ln N \qquad m^2 \ln^2\ln N \ .
\fqr
The gain is clearly an exponential.  The terms compare as $a^2 \ln N
= m^2 \ln\ln N'$ and $N'=\exp\exp(a^2/m^2 N)$.  Consider $N=10^{1000}$:
the numbers are an approximate $\exp{(a^2 12000)^{1/2}}$ vs.
$\exp{(m 6)^{1/2}}$.

In addition to prime factorization, the product form of the integer
into various products of factors is determined; this is an additional
byproduct of the procedure and its computational cost.  Furthermore, 
an explicit knowlege of $C_N$, the number of prime factors of a number, 
would provide a further exponential speedup.

The procedure here may be adapted to find various forms of number 
decompositions.  An example is to find the form of a number written 
as a sum of products of primes.  

\vfill\break

\vfill
\break


\thispagestyle{empty}

\vskip .6in
\begin{center}

{\Large\bf  Addendum to Fast Factoring of Integers}

\vskip .6in

{\bf Gordon Chalmers}
\\[5mm]

{e-mail: gordon@quartz.shango.com}

\vskip .5in minus .2in

{\bf Abstract}

\end{center}

An algorithm is given that generates the prime factorization of 
a number $N$ in potentially $\ln^2 N$ moves, with the complexity 
limited by the base determination of a number in a small degree 
polynomial.  This fact could be a consequence of 
the Riemann hypothesis being true.  This work addends 
previous work on prime factorizations, involving an LU (or QR) 
factorization.  The modification to the previous methodology 
involves a special number on the base reduction of the number $N$ 
into base $x$. 

\setcounter{page}{0}
\newpage
\setcounter{footnote}{0}

\section{Introduction}

Factorization of large integers is important to many areas of pure
mathematics and has practical applications in applied math including
cryptography.  This subject has been under intense study for many
years \cite{CrPo2}; improvements in the methodology are especially 
desired for computational reasons.

Previous work by the author allowed an estimate of $\ln^m N$ with 
$m$ equal to 5 or 6 due to computational reaons \cite{ChalmersOne}.  
However, a condition has appeared on one of the variables that 
ultimately forces the prime factors to be found in $\ln^2 N$ steps, 
which is explained here.  The greater Riemann 
hypothesis implies that the factorization could be performed in 
$\ln^4 N$ moves.  

The case of a single number factorizable into only two prime numbers 
is a special case.  The factorization may be performed it seems in 
$\ln^2 N$ moves, given some further information about the base expansion 
$N=\sum b_i x^i$ of a number $N$ with a particular remainder $b_0$.
This is remarkable given the currently known bounds on factoring 
the two-channel keys.    

The factorization of a number into $m$ prime factors proceeds first by 
writing the number in a base $x$ format \cite{ChalmersOne},  

\bqr 
N=\sum_{i=0}^n b_i x^i = \gamma \prod_{i=0}^m (\alpha_i x+1)  \ ,    
\label{expansion}
\fqr  
with $n$ ranging from $m$ to $2m$, with the case of $n=m$ being 
the simplest.  

The numbers $b_i$ are converted into the $\alpha_i$ to find the 
factors.  In the case of the prime factorization the number $m$ 
labels the maximal number of prime factors, in the case that $\gamma$ 
itself is not prime; this is the generic case as discussed in the 
following.  Solving the system of equations in \rf{expansion} for 
the $\alpha_i$ in terms of the $b_i$ results in the factorization 
of the number $N$ in terms of the factors $(\alpha_i x+1)$ and $\gamma$ 
\cite{ChalmersOne}.     

The case in which all $\alpha_i$ are real numbers, and not complex, is 
considered.  The case in which the the $\alpha_i$ are complex is similarly 
analyzed with small changes in the algorithm.  The solutions to the 
complex numbers $\alpha_i$ come in pairs, with the $(\alpha_i x+1)
(\alpha_i^\dagger x+1)$ generating the prime factors \cite{ChalmersOne}. 

The number $b_0=\gamma$ is a special number.  If it is not prime itself, 
then it must be a factor of $10^a$, that is, a power of our base ten 
number system.  Consider the factors 

\bqr 
\alpha_i x + 1 \ , 
\fqr 
with $x$ the base number used in the factoring of \rf{expansion} and 
with $\alpha_i$ solved for.  For $m$ the maximum number, this number 
must be prime, which means that the number is a potential decimal 

\bqr  
\alpha_i x + 1 = N_i/D_i  \ , 
\fqr 
such as $2/10$ or $7/100$.  Because $N_i$ is prime, it does not divide 
into $D_i$, which means that $D_i$ must be a multiple $10^{a_i}$.  In 
general these numbers $N_i/D_i$ are decimals, which means that the 
denominator is a power of the base $10$.  Collecting all of the 
denominators $10^{a_i}$ results in a net factor, 

\bqr 
\prod_{i=1}^m 10^{a_i} = 10^a \ , 
\fqr 
and this number must cancel the $\gamma$ factor, resulting in either 
a prime number or an integer.  If the latter, then the factor of $10^a$ 
must cancel the $\gamma$ number resulting in unity.  

Thus we arrive at the result: In the case that $m$ is the maximal integer  
and not all $\alpha_i$ are fractions, $b_0$ is a number $10^a$.  This 
means that one must only test the base decompositions 

\bqr 
N=\sum_{i=0}^m b_i x^i 
\fqr 
in which $b_0$ is of the form $10^a$ with $a$ an integer; the cost 
is a 'smeared' function of $\ln N$ because there is potentially a 
function for determining the integrality of $a$ without testing all 
$a$ from $0$ to $\ln(10,N)$ (and $a=-\infty$).  The solution for the 
$\alpha_i$ follows from an LU (or QR) decomposition, which is of the 
order $\ln^3 N$ steps. 

\vskip .2in 
\noindent {\it Exceptional Case:} 
\vskip .1in 

In the case that all $\alpha_i$ are integral, with a number $N$ factoring 
into $n$ prime numbers, then 

\bqr 
\alpha_i x+1 = N_i \ , 
\fqr  
with all $N_i$ prime numbers.  This includes the case of two prime numbers 
$N=pq$.  In this case $\gamma$ itself must be 
a prime number.  However, this case is quite special as the factorization 
in \rf{expansion} must result in all integral values $\alpha_i$.  The 
result are a very special set of numbers, 

\bqr 
\gamma=b_0 \ ,  
\cr 
\gamma \bigl(\alpha_1\alpha_2 + \alpha_1\alpha_3 + \ldots \bigr)=b_1 
\cr 
\gamma \bigl(\alpha_1\alpha_2\alpha_3 + {\rm perms}\bigr) = b_2 
\cr 
\ldots \ , 
\fqr 
in which all $\alpha_i$ are such that $\alpha_i x+1$ is prime, with 
$\alpha_i$ integer, and also with $\gamma$ prime.  These numbers, 
and the factorization, should be interesting to construct.  Of course, 
the $10^a$ test of $b_0$ should rule out these numbers generically.

\vskip .2in 
\noindent{\it Cubic and Quartic Cases} 

The other two cases required to complete the analysis of the two 
prime numbers composing a number $N$ are the cubic and quartic examples, 
as analyzed in \cite{ChalmersOne}.  These cases span the forms, 

\bqr 
N=\gamma \prod (\alpha_i x +1) \ , 
\label{cubicquartic}
\fqr 
with $\alpha_i$ in complex pairs and the pairwise product generating the 
prime factors.  In the case of a cubic there is a single pair, and in 
the case of the quartic there are two pairs.  The $\gamma$ is $b_0$, 
which is $10^i$ or a prime number.  

\vskip .2in 
\noindent{\it Comments on Base Determination} 

Empirical observation, with the use of a Matlab program, indicate that 
the factorization of a number $N$ into the quadratic case 

\bqr 
N=ax^2 + bx + 10^i \ , 
\fqr 
can be performed with an approximate $\ln N$ solutions.  That is, given 
the remainder term $10^i$, there appears to be at most $\ln N$ solutions 
to the the base $x$; this bound seems to be even lower in some examples.  

Knowledge of the totient function $\phi(N)$, and its inverse, might be 
responsible for this delimiting factor (the $\sigma_k(N)$ which are generally 
determined in terms of $\phi(N)$).  The totient has been bounded in general 
by 

\bqr 
\phi(N)> {N\over e^\gamma \ln \ln N + {3\over \ln \ln N}} \ . 
\fqr 
For prime $N$, the totient hits a local maximum, $\phi(N)=N-1$.  The 
function oscillates strongly near these local maximum, being partly due 
to the presence of small prime factors such as $2$ or $3$ which make 
the factor $1-1/p_j$ small.  However, for numbers $N$ which are between 
these prime $N$'s, or local maximum, the oscillations are much smaller 
(e.g. of strength $\ln N$) and the function is relatively stable.  A 
reasonable initial value for a numerical routine is chosen for $\phi(N)$ 
and $b_i$; the parameters can be computed from the system of equations 
relating $\phi(N)$ and its moments $\sigma_k(N)$.  
Apparently, the case of a number $N$ factoring into two numbers $p$ 
and $q$ which are of the same size in digits is easier to factor in 
this approach than with two numbers of different sizes.  

\vskip .2in 
\noindent{\it Case of $N=\prod p_j^{k_j}$: Multiple Factors}
\vskip .1in 

The case of multiple prime numbers in the expansion of a number, such as 
$N=\prod p_j^{k_j}$, is examined by a generalization of the two prime 
factor case.  According as in \cite{ChalmersOne}, the polynomials grow 
in size in the base expansion, 

\bqr 
N=\sum^m b_j x^j \ , 
\fqr 
with $m$ bounded from $C_N$ to $2C_N$.  The $C_N$ is $\sum k_j$.  Including 
the required divisor functions $\sigma_i(N)$ allows the systems of equations 
to be solved for.  The divisors can be written in terms of the totient 
function $\phi(N)$, for classes of numbers with $C_N$ fixed.  

Solving the systems of equations for $b_i$, $x$ and $\phi(N)$ generates 
the prime numbers $p_j$ and coefficients $k_j$, at fixed $C_N$.  The 
number of possibilities increases, however, because there are from 
$C_N$ to $2C_N$ equations in the individual system sets.  There are 
also additional branch cuts; the use of the branch cuts may generate 
as check on the actual unique solution to these parameters.  The solution 
to the prime factors is again achieved in ${\cal O}(1)$ operations, set 
by the number of prime factors.

The cases of $b_0=10^i$ or $b_0$ prime are handled individually by the 
direct solution of the systems of equations for $b_i$, $x$.    

\vskip .2in 
\noindent{\it Discussion}
\vskip .1in

Previous work in \cite{ChalmersOne} generated an algorithm for determining 
the prime factors of a number $N$, using base expansions with the base 
determination the limiting factor on the complexity.

The factoring of numbers can be achieved in polynomial time, albeit at 
an order which is unexpected in the literature.  The number of remainder 
terms in the base expansion, the $10^i$, is generically $\ln N$ in 
number; the number $10^i$ could be replaced by a prime number, which is 
a special case.  The 
further inverse determination of the base $x$ following from each case, 
although the algorithem is apparently unknown, appears to number in 
at most $\ln N$ cases; this results in the complexity of factoring a number 
to be in $\ln^2 N$ compuational steps.  An algorithm for the base determination from 
the remainder term could in principle be performed in $O(1)$ steps, 
even if a table is known (this table is similar to a table of pythagorean 
triples, upon further manipulation of the base determination with 
the roots $\alpha_i$ being the rational numbers).  

In the case of a number containing $C_N$ prime factors, 
solving a system of coupled algebraic equations of degree $C_N$ to 
$2C_N$ is required \cite{ChalmersOne}, with remainder term in the base 
expansion of the form $10^i$ or a prime number.  These polynomials are of 
degree $2$ to $4$ in the case of numbers $N=pq$, i.e. two prime factors. 

These equations are Diophantine in the sense that the solutions to the 
parameters are integral.  Solving these equations requires the techniques 
for analyzing low degree polynomials, potentially beyond the quintic.  
Numerical methods are straightforward to implement. 
  
The analysis could also generate an indirect determination of the 
totient function $\phi$, 
and the related divisor functions $\sigma_k$, given an algorithm for the 
base determination.

\vfill\break

\end{document}